\documentclass[aps,prd,reprint,superscriptaddress,nofootinbib,showkeys]{revtex4-2}
\usepackage[T1]{fontenc}
\usepackage[utf8]{inputenc}
\usepackage{lmodern}
\usepackage{amsmath,amssymb}
\usepackage{graphicx}
\usepackage{xcolor}
\usepackage{pgfplots}
\usepackage{microtype}
\usepackage[unicode,hidelinks]{hyperref}
\usepackage{url}
\usepackage{booktabs}

\begin{document}

\title{{Two-Photon Interactions at High Energies: Establishing a Theoretical Benchmark for the Color Dipole Formalism and Saturation Models}}
\author{Zardo Becker, G.}
\affiliation{Physics Institute, Federal University of Santa Catarina (UFSC), Florianópolis/SC, Brazil}
\affiliation{Physics Institute, The Ohio State University (OSU), Columbus/OHIO, United States}
\email{zardobecker@gmail.com}
\author{Benno Bodmann}
\affiliation{International Center for Relativistic Astrophysics Network headquartered on Universidade Federal do Rio Grande do Sul (ICRANet/UFRGS) , Porto Alegre, Brazil}
\email{benno.bodmann@gmx.de}
\author{C\'esar A. Zen Vasconcellos}
\email{cesaraugustozenvasconcellos@gmail.com}
\affiliation{International Center for Relativistic Astrophysics Network (ICRANet), Pescara, Italy}

\begin{abstract}
{The transition of the photon from an effectively point-like structure to hadron fluctuations at high energies requires a rigorous description of gluon dynamics in two-photon collisions. To map this transition without resorting to purely empirical artifices, it becomes necessary to establish a controlled analytical baseline — a benchmark — that examines the foundations and operational limits of the color-dipole framework. Because standard low-energy frameworks like the Quark-Parton Model and Vector Meson Dominance cannot accommodate the gluon component demanded by LEP results, we must evaluate the foundational limitations of the TKM and IKT prescriptions. We explicitly track their structural compromises—including exchange symmetry breaking, the homogeneous-disk target approximation, and the hypothesis of universality—to understand where effective parameterizations substitute for first-principles of QCD. By articulating the non-linear Balitsky--Kovchegov evolution equation with five saturation models — GBW, IIMS, rcBK, AGBS, and WYKWC — the study maps the theoretical dispersion arising from different treatments of the saturation scale and coupling. This methodological reference aims to dimension systematic uncertainties, test the consistency of the dipole-proton analogy, and subsidize data analysis in future electron-positron colliders, as treated in detail in the companion paper \citep{ZardoBecker2026b}.}
\end{abstract}


\maketitle
\tableofcontents

\section{Introduction}

{When particles collide at high energies with large momentum transfer, their internal parton constituents are spatially and momentum-resolved by the probing current. While deep inelastic scattering has long mapped this parton interior in protons, the photon—traditionally viewed as an elementary, structureless gauge boson—exhibits a rich hadron substructure when probed at sufficiently high energies, fluctuating into quark-anti-quark pairs and multi-gluon states through its Fock-space decomposition~\citep{Brodsky:2006,Nisius:2000,Budnev:1975}:
\[
|\gamma\rangle = |\gamma_0\rangle + |q\bar{q}\rangle + |q\bar{q}g\rangle + \cdots.
\]
Electron-positron colliders provide a pristine laboratory to study this transition via bremsstrahlung interactions, allowing us to access hadron production in $\gamma\gamma$ collisions. While low-energy interactions are successfully described by the Quark-Parton Model~\citep{Gotsman:1988,Gotsman:2000} and Vector Meson Dominance~\citep{Donnachie:2000}—where the photon behaves as an effective superposition of vector bound states or free constituent quarks—these frameworks encounter a fundamental breakdown as the center-of-mass energy increases. The observed rapid rise of the cross section at LEP energies cannot be accommodated by soft Reggeon exchanges or finite-order quark boxes because they omit the dense gluon cascading that governs the high-energy asymptotic regime. Accounting for this growth requires incorporating gluonic degrees of freedom through the color-dipole formalism~\citep{Abbiendi:2000a,Timneanu:2002}. However, because a full QCD derivation of the scattering between two extended, fluctuating quantum systems remains mathematically and computationally intractable, the theoretical description must rely on effective prescriptions.}

{To bridge this analytical gap, the literature adopts the TKM~\citep{Timneanu:2002} and IKT~\citep{Iancu:2008} frameworks, which map the mutual scattering of extended color dipoles onto the established dipole-proton amplitude $\mathcal{N}(r,Y)$ extracted from deep inelastic scattering. Rather than treating these prescriptions as exact solutions, we explicitly quantify their underlying structural compromises: when interacting dipoles possess comparable transverse sizes, both models artificially break the $1 \leftrightarrow 2$ exchange symmetry, forcing an arbitrary designation of projectile and target; they model the target dipole as a homogeneous disk, thereby discarding impact-parameter dependence; and they invoke an unproven hypothesis of universality by transferring the proton-target amplitude directly to photon-photon collisions without dynamical re-adjustment. }

{
To understand what this divergence implies physically, consider how each prescription treats the geometry of the collision. The TKM approach forces the two interacting dipoles into a symmetric effective size $r_{\text{eff}} = r_1 r_2 / \sqrt{r_1^2 + r_2^2}$, treating both objects on equal footing regardless of their individual dimensions. In contrast, the IKT approach enforces an asymmetric hierarchy: it treats the smaller dipole as a quantum projectile probing a static target represented by the larger dipole, bounded by an arbitrary momentum cutoff $\Lambda_c \sim \Lambda_{\text{QCD}}$. When the two dipoles have comparable sizes, these two opposing geometric assumptions pull the cross-section calculations in opposite directions. Specifically, TKM expands the effective interaction region by treating both objects symmetrically—{allowing large fluctuations from both fields to constructively enhance the mutual overlap}—whereas IKT restricts the interaction via an asymmetric momentum cutoff that shrinks the effective cross section as energy grows—{artificially freezing the target's internal dynamics and imposing an upper limit on the transverse momenta that can contribute to the scattering}. The resulting systematic divergence is not merely a numerical discrepancy; it is the visible mathematical footprint of two fundamentally different ways of idealizing a complex, non-perturbative collision, defining the precise boundary where our current theoretical models lose their first-principles footing.}

{The present paper, the first of two companion works, establishes a controlled theoretical baseline designed to examine these foundational limitations. By systematically articulating the non-linear Balitsky--Kovchegov evolution equation~\citep{Kovchegov:1999,Kovchegov:2000} with five distinct saturation models—GBW~\citep{GolecBiernat:1998}, IIMS~\citep{Iancu:2004}, rcBK~\citep{Balitsky:1996,Kovchegov:2007,Albacete:2009}, AGBS~\citep{Amaral:2007}, and WYKWC~\citep{Wang:2021}—we construct a rigorous theoretical envelope that maps the dispersion arising from different treatments of the saturation scale and running coupling. This methodological framework provides a quantitative baseline to test the limits of the dipole-proton analogy, establish falsifiable criteria for future precision measurements at high-energy linear and circular colliders (such as the ILC, CLIC, and FCC-ee), and isolate where effective parameterizations must eventually yield to a complete, first-principles QCD treatment, as detailed in the companion paper~\citep{ZardoBecker2026b}.}

To ensure that the TKM--IKT ambiguity is not merely a theoretical artefact but a falsifiable hypothesis, we establish the following criterion: if future measurements yield the total photon-photon cross-section $\sigma_{\gamma\gamma}(W, Q^2=0)$—{evaluated strictly in the real-photon limit where the probing virtuality vanishes and fluctuations into extended hadron states are fully unsuppressed, maximizing the sensitivity to parton substructure}—with experimental errors comparable to the typical precision of these colliders (expected to be below the 20\% level for inclusive cross sections) and the central values fall within the overlapping bands of TKM and IKT predictions at $W \sim 1$ TeV, the hypothesis of a systematically growing divergence between the two prescriptions will be considered refuted. {Operationally, this implies that if precision data cannot separate TKM's symmetric effective size $r_{\text{eff}}$ from IKT's asymmetric momentum cutoff $\Lambda_c$, it means inclusive measurements alone are blind to the geometric competition between expansion and restriction, rendering the theoretical differences experimentally indistinguishable at these energy scales and demanding more exclusive, differential observables to break the tie.} Such a result would indicate that the theoretical distinction between the two frameworks remains experimentally inaccessible, implying that more sensitive observables will be required to resolve the underlying ambiguity. Conversely, if the data lie outside the combined envelope of the five saturation models, neither prescription will be able to describe the data, signaling a breakdown of the dipole--proton analogy itself. {Because these five models comprehensively cover the theoretical spread of known gluon saturation mechanisms—spanning from coordinate-space parameterizations to running-coupling evolution—falling outside their collective prediction band means that simply tweaking the saturation scale or running coupling is mathematically futile.} Such an extreme discrepancy would demonstrate that adjusting the non-linear BK evolution or the saturation scale is no longer sufficient; it would expose a foundational failure in assuming that a target amplitude extracted from deep inelastic scattering on protons can be transferred without dynamical modification to the mutual collision of two fluctuating photon fields. Such a discrepancy would point to uncaptured higher-order fluctuations or novel non-perturbative dynamics that fundamentally invalidate the foundational assumption of mapping photon-photon scattering onto standard dipole--proton frameworks.

With the prospect of future high-energy electron--positron colliders, it will become possible to probe hadron production from $\gamma^{(*)}\gamma^{(*)}$ interactions at unprecedented energies. These measurements will help reduce theoretical uncertainties and enhance the precision of QCD-based predictions. While one of the main objectives of these colliders is the study of the Higgs boson~\citep{Asner:2003, dEnterria:2017} and the search for signals of physics Beyond the Standard Model (BSM)~\citep{Boyko:2023}, many of the relevant final states involve hadrons. Since hadron production in two-photon interactions is directly related to the parton density in the photon, it becomes essential to analyze the behavior of different dipole scattering amplitude models and dipole-dipole cross section formulations. This understanding is crucial for accurately modeling and estimating the hadronic background in precision measurements at future colliders.

The content of this paper is organized as follows. Sec.~\ref{sec:two_photon} presents the three main contributions to the total $\gamma^{(\ast)}\gamma^{(\ast)}$ cross section: VDM, the QPM and the gluon contribution (dipole formalism). The latter is modeled using two different prescriptions for the dipole–dipole cross section, namely TKM and IKT, which are described in detail in Sec.~\ref{sec:dipole-dipole}. These prescriptions depend on the dipole scattering amplitude, governed by the non-linear BK evolution equation. The BK-based models used in this work - rcBK, GBW, IIMS, AGBS, and WYKWC - are discussed in Sec.~\ref{sec:BK_models}, where their phenomenological saturation features are also analyzed. The companion paper~\citep{ZardoBecker2026b} presents the results of applying these models to physical observables, quantifies the TKM–IKT ambiguity, tests universality, estimates fragmentation contributions, and identifies experimental discriminants for future colliders.

\section{Two photon interaction}\label{sec:two_photon}

\begin{figure}[htb]
    \centering
    \includegraphics[width=0.8\linewidth]{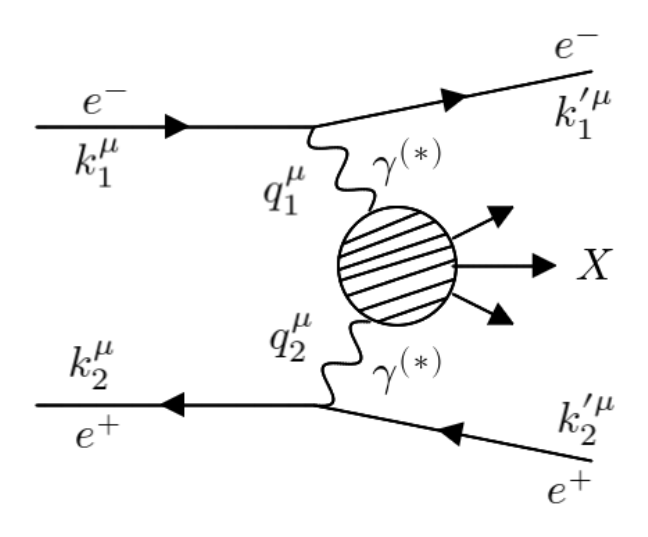}
    \vspace{-0.3cm}
    \caption{Pictorial illustration of electron-positron scattering with two-photon interaction producing hadrons in final state ($X$), where the $q^\mu_i = k^\mu_i - k^{\prime\mu}_i$ is photon momentum writed by the difference of initial ($k^\mu_i$) and final ($k^{\prime\mu}_i$) of the lepton momentum.}
    \label{fig:eletron-positron}
\end{figure}

The production of hadrons in electron–positron elastic scattering can be explained by photon–photon interactions influenced by QCD dynamics, as illustrated in shaded area of Fig.(\ref{fig:eletron-positron}). Several models of two-photon interactions have been developed to describe hadron production, taking into account the varying dynamics depending on the photon virtuality \citep{Nisius:2000}. Most of these models combine the VDM approach with the QPM, suitably extended into the region of low virtuality, which dominate at low energies. However, to describe the observed growth of the cross section with increasing energy in experimental data, it becomes essential to incorporate QCD dynamics. In particular, the gluon-driven contributions become increasingly important, similar to what is predicted in photon–proton scattering.

We use the standard notation: $Q^2_i = -q_i^2$ is the photon virtuality, $W^2_{\gamma \gamma} = (q_1 + q_ 2)^2$ is the $\gamma^{(\ast)}\gamma^{(\ast)}$ c.m. energy. In these two-photon reactions, the photon virtuality can be made large enough to ensure the applicability of perturbative methods or can be varied in order to test the transition between the soft and hard regimes of QCD dynamics.
The photon can act either as a structureless particle or manifest as a quark–anti-quark pair through a perturbative fluctuation. Examining the individual contributions to the total cross section provides valuable insight into how short-range, high-energy interactions transition into long-range, non-perturbative effects.

\subsection{Vector Meson Dominance (VDM)}

The non-perturbative contribution from the coupling of Reggeons to the hadronic fluctuations of the photon carries inherent ambiguities concerning the formation of meson states in the low-energy regime~\citep{Donnachie:2000}. Even when one photon is real, corresponding to the valence-quark component, significant uncertainties persist regarding the effective coupling and the transition to hadron states.
In the Vector Meson Dominance (VMD) approach, we therefore adopt a framework in which the hadron part of the photon is treated as a superposition of quark–anti-quark dipoles, rather than being explicitly modeled as vector mesons.

As described in~\citep{Donnachie:2000}, a simple approximation for the case in which both photons fluctuate into vector meson states allows the process to be treated as a contribution from Reggeon exchange, which can be factorized in the form:
\begin{equation}\label{eq.Sig_reg1}
\begin{aligned}
\sigma_{Reg}^{\gamma^{(\ast)} \gamma^{(\ast)}}(W^2, Q_{1,2}^2) 
\! = \! \pi^2 \alpha_{em}^2 \frac{C}{a} \left(\frac{W^2}{a}\right)^{\!\!-\eta} 
\!\! \Biggl[\frac{a^2}{(a+Q_1^2)(a+Q_2^2)}\Biggr]^{1-\eta},
\end{aligned}
\end{equation}
{where the symbol $\sigma_{Reg}^{\gamma^{(\ast)} \gamma^{(\ast)}}(W^2, Q_{1,2}^2)$ denotes the soft non-perturbative cross section for the collision of two photons—which can be real ($\gamma$) or virtual ($\gamma^*$)—characterized jointly by their invariant center-of-mass energy squared $W^2$ and their individual photon virtualities $Q_1^2$ and $Q_2^2$.} Here, $\alpha_{em}$ is the electromagnetic coupling and the parameters $\eta = 0.3$ and $C = 0.26$ are associated with the non-perturbative contributions of Reggeon exchange {—representing physically the soft, long-range background where the photon acts as a diffuse cloud of light vector mesons undergoing smooth Regge trajectories}. In this parametric structure, the term $(W^2/a)^{-\eta}$ encodes the characteristic power-law decline of soft hadron exchanges with rising center-of-mass energy $W$, while the squared bracket containing the virtualities $Q_{1,2}^2$ acts as a multi-pole form factor that suppresses the cross section as the colliding photons depart from the real-photon limit and become increasingly point-like and off-shell. The coherence length is related with parameter \linebreak $a = 0.2~\text{GeV}^2$, were fitted with data on photon-photon collision to incorporates aspects of saturation physics \citep{Timneanu:2002}. This formulation governs the energy dependence by suppressing soft Reggeon dominance at high energies through a reduced coherence length {—acting effectively as an inverse-range mass scale that prevents soft interactions from overrunning the cross section}. The parameters fitted in \citep{Timneanu:2002} yield lower values than the original expression of \citep{Donnachie:2000}, which was restricted to low-energy kinematics ($W < 10~\text{GeV}$) {—reflecting the mathematical necessity to retune the soft background so it does not falsely overshoot the explosive growth observed at LEP energies before non-linear gluon dynamics take over}.

The decomposition of the Reggeon contribution according to photon polarization states remains undefined in a rigorous sense. To proceed with our analysis, we assume  that transversely polarized photons is the domaine state of the Reggeon couples. Under this assumption, the full expression for the Reggeon contribution, including the correction term to smooth the contribution of Reggeons to small x , with the Eq.~(\ref{eq.Sig_reg1}) the VDM cross section is:
\begin{equation}\label{eq.Sig_reg}
\sigma_{R}^{\gamma^{(\ast)} \gamma^{(\ast)}}(W^2, Q_{1,2}^2) = \sigma_{Reg}^{\gamma^{(\ast)} \gamma^{(\ast)}}(W^2,Q_{1,2}^2)(1-\bar{x}) \ ,
\end{equation}
where $\bar{x}$ is the modified Bjorken-$x$ variable used for two photons as
\begin{equation}
\bar{x} = \frac{Q_1^2 + Q^2_2 + 8m^2_q}{W^2 + Q_1^2 + Q^2_1} \ .
\end{equation}

\begin{figure}[htbp]
    \centering
    \includegraphics[width=1.0\linewidth]{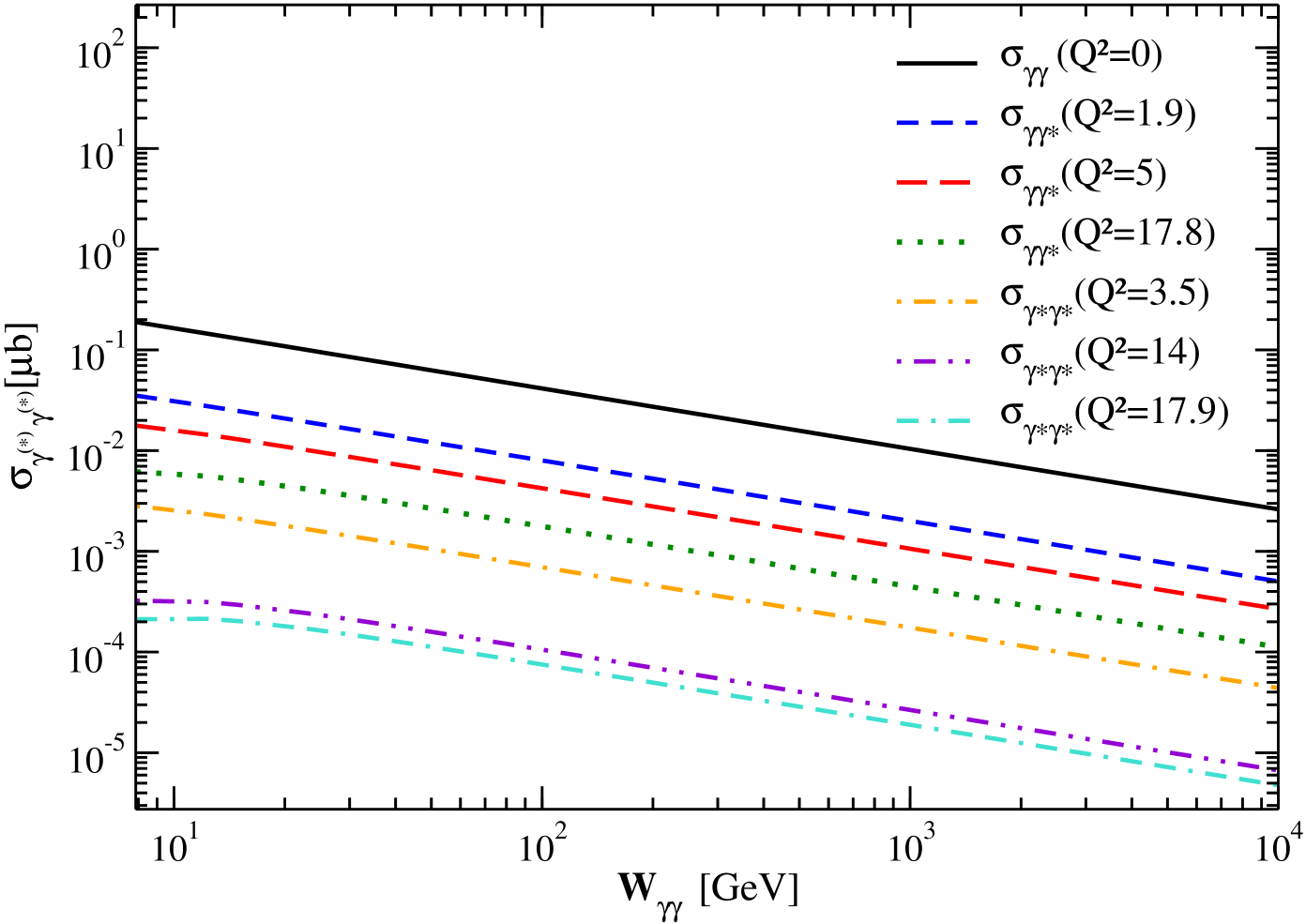}
    \caption{Theoretical result of Reggeons exchange contribution for cross section of: two real photon $\sigma^{\gamma\gamma}$ with $Q^2 = Q_1^2 = Q^2_2 \approx 0$; two virtual photon $\sigma^{\gamma^\ast \gamma^\ast}$ with $Q^2 = Q_1^2 = Q^2_2$, and real-virtual $\sigma^{\gamma \gamma^\ast}$ with $Q^2_1 \approx 0$ and $Q^2 = Q^2_2$. Analyses of different virtuality, show how the Reggeons contribution with c.m. energy of photon-photon interaction.}
    \label{fig:Reggeons_result}
\end{figure}

Equation~(\ref{eq.Sig_reg}) exhibits explicit dependence on the light-quark masses $m_q = m_{u,d,s}$. Figure~\ref{fig:Reggeons_result} displays the corresponding numerical results as a function of the center-of-mass energy across various virtuality, utilizing $m_{u,d,s} = 0.210~\text{GeV}$ to illustrate the baseline behavior of the Reggeon-exchange contribution.

The figure shows a soft decrease in the cross section with increasing energy, as well as a further reduction with increasing photon virtuality. This results in a very small contribution when both photons are virtual and a bigger contribution when at least one real photon is present in the interaction. {Physically, this occurs because increasing virtuality ($Q^2$) shrinks the photon's transverse size and suppresses its extended hadron cloud, turning it into a point-like probe that rarely interacts, whereas real or quasi-real photons ($\sim 0$) maximize spatial fluctuations and allow the soft hadron mechanisms to operate fully.}

The ambiguity in the polarization decomposition of the Reggeon contribution, together with the adjustment of parameters to describe the transition to higher energies, illustrates one of the several sources of phenomenological uncertainty that our benchmark aims to quantify.

\subsection{Quark Parton Model (QPM)}

The production of hadrons from two photons at low energy was extensively discussed by~\citep{Budnev:1975}. The approach used by the author to describe hadron production in $\gamma^{(\ast)}\gamma^{(\ast)}$ interactions relies on the Quark-Parton Model (QPM). It can be expressed in terms of the two-photon cross section for lepton production, as given in~\citep{Budnev:1975}: 
\begin{equation} \label{eq.QPM1}
\hspace{-0.2cm}
\sigma_{QPM}({\gamma^{(\ast)} \gamma^{(\ast)}} \rightarrow X) =  \sum_f e_f^4\sigma_f(\gamma^{(\ast)} \gamma^{(\ast)} \rightarrow \mu^-\mu^+) \ ,
\hspace{-0.2cm}
\end{equation}
where $e_f$ is the electric charge of a quark of flavor $f$. {Physically, this foundational relation dictates that quark-anti-quark production is calculated by scaling the point-like QED cross section for muon pairs with the respective fractional charges $e_f^4$, treating quarks as quasi-free point-like constituents without internal binding dynamics.} Eq.~(\ref{eq.QPM1}) is valid in the limit $W^2 \gg |q_i^2|$, where $q_i$ is the four-momentum of photon $i=1,2$. 
\begin{figure}[htbp]
    \centering
    \includegraphics[width=0.9\linewidth]{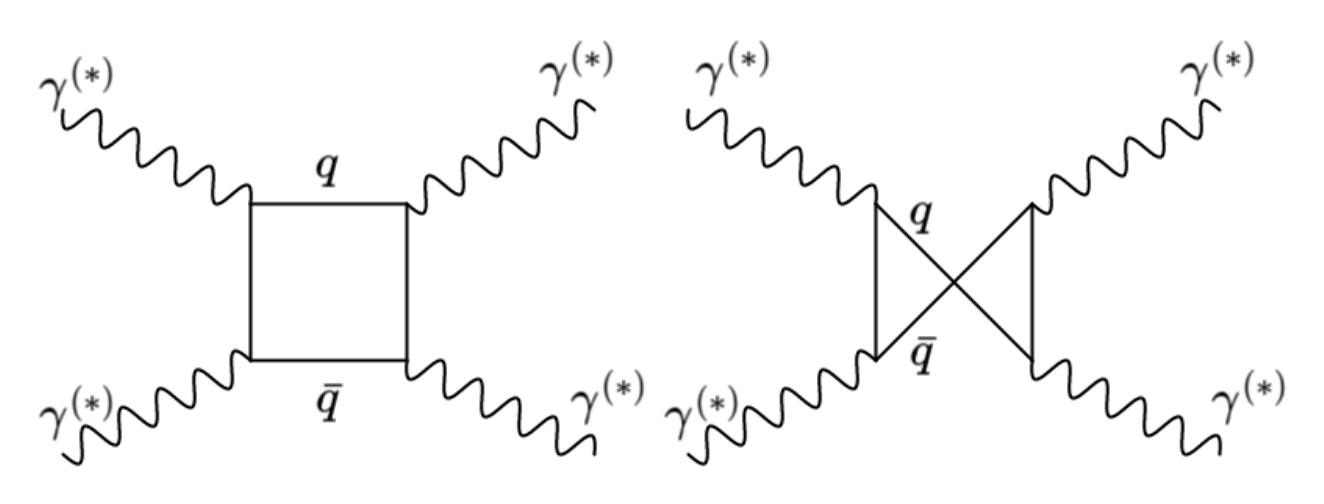}
    \caption{Box diagram for the quark parton model contribution to the total photon-photon cross section.}
    \label{fig:QPM}
\end{figure}
This is the only reaction in which the masses of the produced particles can vary over a wide range. As a result, different particles can be produced in these reactions, such as various types of hadrons in the final state, which are not directly \linebreak observed but influence the photon scattering process. Thus, the $\gamma\gamma$ cross section is correlated with the forward $\gamma\gamma$ scattering amplitude via unitarity and optical theorems \citep{Budnev:1975}, {linking the inclusive production of unobserved multi-hadron states directly to the imaginary part of the forward elastic scattering amplitude}

\begin{subequations}\label{eq.QPM}
\begin{align}
\sigma_{TT}^{\gamma^{(\ast)} \gamma^{(\ast)}} 
&= \frac{\pi \alpha_{em}^2}{W^2 x}
\Biggl\{ 
q_1 q_2 L \Biggl[ 2 + \frac{2m^2}{x} - \left(\frac{2m^2}{q_1 q_2}\right)^2 \notag \\
&\qquad + \frac{q_1^2 + q_2^2}{x} 
  + \frac{q_1^2 q_2^2 W^2}{2x (q_1 q_2)^2} 
  + \frac{3}{4} \left(\frac{q_1^2 q_2^2}{x q_1 q_2}\right)^2 \Biggr] \notag \\
&\qquad - \Delta t \Biggl[ 1 + \frac{m^2}{x} + \frac{q_1^2 + q_2^2}{x} 
  + \frac{q_1^2 q_2^2}{T} + \frac{3}{4} \frac{q_1^2 q_2^2}{x^2} \Biggr]
\Biggr\}, \\[4pt]
\sigma_{LL}^{\gamma^{(\ast)} \gamma^{(\ast)}} 
&\!\!\!=\! \frac{\pi \alpha_{em}^2 q_1^2 q_2^2}{W^2 x^3}  \!
\Biggl\{\!\! \frac{L}{q_1 q_2} \bigl( 2 W^2 x + 3 q_1^2 q_2^2 \bigr) 
      - \Delta t \left(\! 2 + \frac{q_1^2 q_2^2}{T} \! \right) \!\!\! \Biggr\}.
\end{align}
\end{subequations}
and mix of polarized states (TL) and (LT) is
\begin{align}
        \sigma_{TL}^{\gamma^{(\ast)} \gamma^{(\ast)}} &= - \frac{\pi \alpha^2_{em} q_2^2}{W^2x^2}\biggl\{
\Delta t \left[1 +\frac{q_1^2}{T}\left( 6m^2 + q_1^2 + \frac{3}{2}\frac{q_1^2 q_2^2}{x}\right) \right] 
\notag \\ \notag & \hspace{1cm}
 -\frac{L}{q_1q_2}\Biggl[ 4m^2x + q^2_1(w^2 + 2m^2)
\notag \\ & \hspace{1cm}
 + q^2_1\left(q_1^2 + q_2^2 + \frac{3}{2}\frac{q_1^2q_2^2}{x} \right) \Biggr]\biggr\} \label{eq.sig_qpm_TL} \ , \tag{5c}
\end{align}
\begin{align}
\sigma_{LT}^{\gamma^{(\ast)} \gamma^{(\ast)}}  &= \sigma_{TL}(q_1^2 \longleftrightarrow q_2^2) \ , \tag{5d} \label{eq.sig_qpm_LT} 
\end{align}
with,
\begin{subequations}
\begin{align}
& W^2x = (q_1q_2)^2 - q_1^2 q_2^2 \ , \\ 
& q_1^2 q_2^2 = (W^2 + Q_1^2  + Q_2^2)/2 \ , \\
& \Delta t  \equiv  t_{max} - t_{min} = \sqrt{4x(W^2 - 4m_f^2)} \ , \\
& T = (m_f^2 - t_{max})(m_f^2 - t_{min}) = 4x m_f^2 + q_1^2q_2^2 \ , \\
& L = \ln\left(\frac{m_f^2 - t_{min}}{m_f^2 - t_{max}}\right) = \ln\left(\frac{(q_1q_2 + \frac{1}{2}\Delta t)^2}{4xm_f^2+ q_1^2 q_2^2} \right) \ ,
\end{align}
\end{subequations}
where $m_f$ is the quark mass with flavor $f$, whose values used in this work follow the updated parameters from \citep{Gotsman:2000} are fixed: $m_{u,d} = 300~\text{MeV}$, \linebreak $m_s = 500~\text{MeV}$, and $m_c = 1500~\text{MeV}$. 

\begin{figure}[htbp]
    \centering
    \includegraphics[width=1.0\linewidth]{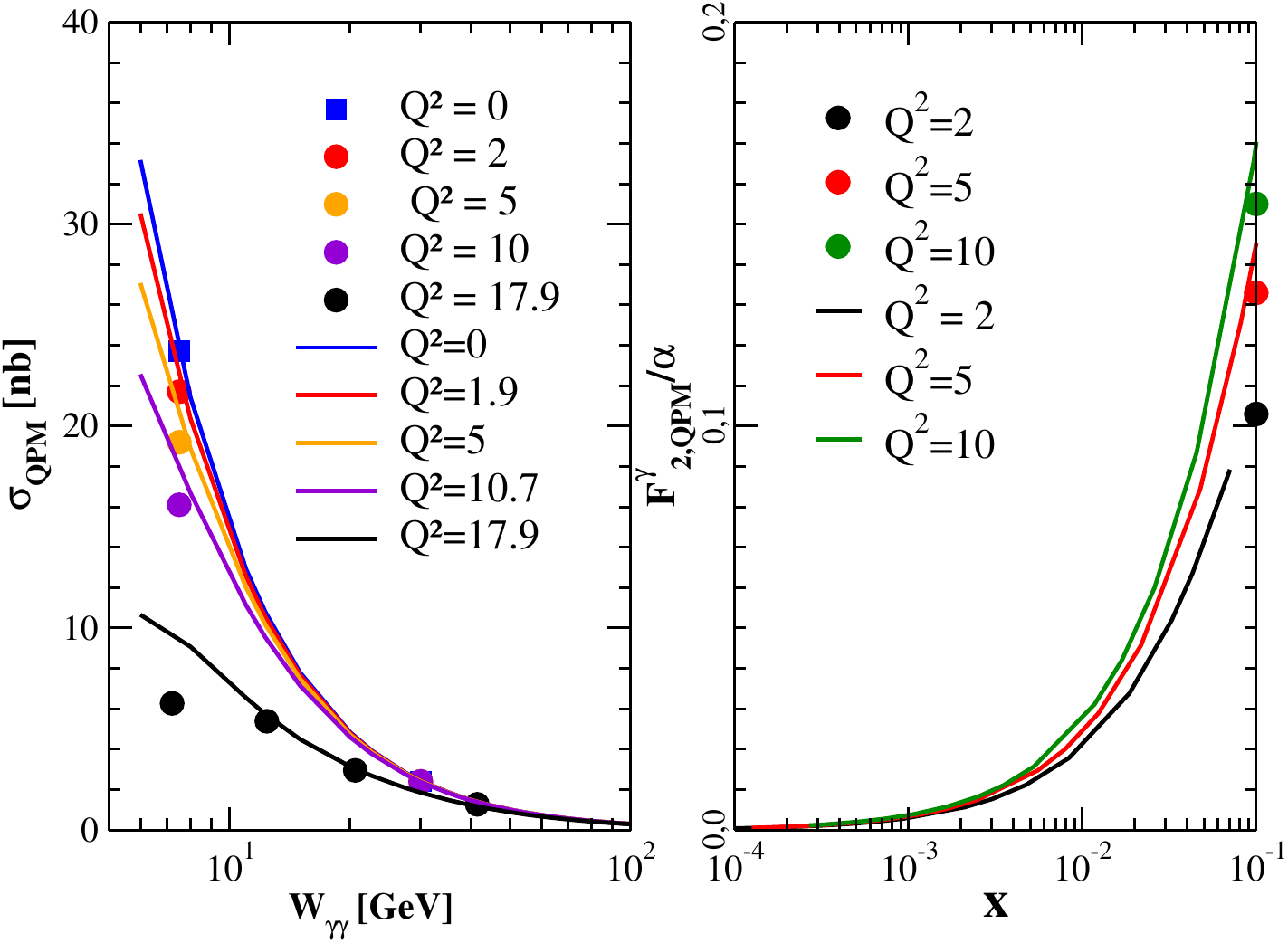}
    \caption{Theoretical quark parton model (QPM) results are compared with experimental data for $x > 0.1$ obtained in (Tab. 1)~\citep{Gotsman:1988} and \citep{Abbiendi:2002b}. In the \textit{left} is shown the real photon ($Q^2=0$) and virtual photon ($Q_1^2 = Q_2^2 = Q^2$) cross section with c.m. energy. In the \textit{right} is show the photon structure function with Bjoken-$x$, when one photon is real ($Q_1^2 \approx 0$) and another is virtual ($Q_2^2 = Q^2$).}
    \label{fig:QPM_results}
\end{figure}

{In summary, what these polarized cross-section formulations ($\sigma_{TT}$, $\sigma_{LL}$, $\sigma_{TL}$, $\sigma_{LT}$) and constituent quark masses reveal physically is that, within the Quark-Parton Model, the photon is treated as an ensemble of quasi-free point-like partons interacting via clean electroweak box diagrams. While this framework successfully captures inclusive scattering data when momentum transfers or constituent masses dominate, it inherently omits non-linear gluon dynamics—laying the exact theoretical foundation and necessity for the color-dipole and saturation formalisms developed next.}

The theoretical results for the photon structure function $F^\gamma_{2,\text{QPM}}  $, when the virtual photon $Q^2_2 \neq 0$ probe the structure of the real ($Q^2_1 \approx 0)$ that is calculated by:
\begin{equation}
F^\gamma_{2,\text{QPM}}(x,Q^2) = \frac{Q^2}{4\pi^2\alpha_{em}} \left( \sigma^{\gamma^\ast \gamma}_{LT, \text{QPM}} + \sigma^{\gamma^\ast \gamma}_{TT,\text{QPM}} \right) \ ,
\end{equation}
and for two photon real(virtual) cross section $\sigma_{QPM}^{\gamma^{(*)} \gamma^{(*)}}$ are presented in Fig.(\ref{fig:QPM_results}). The theoretical result provides a good description of the experimental data and explicitly shows the decrease in the cross section with increasing energy for different photon virtuality. The QPM contribution becomes small for $W > 100~\text{GeV}$ for $\sigma_{QPM}^{\gamma^{(*)} \gamma^{(*)}}$ (\textit{left)} and $x < 10^{-3}$ for $F_{2,\text{QPM}}$ (\textit{right}).

The QPM description, while successful at low energies, fails to account for the rising cross section observed at LEP energies (see Fig.~\ref{fig:QPM_results}). This failure motivates the gluon contribution presented in the next subsection, and underscores the need for a theoretical benchmark that systematically maps the transition from the soft to the hard regime.

\subsection{Gluonic contribution}

The photon-photon interaction can be naturally explained in the language of dipoles, considering that the interaction between photons is directly related to the dipole-dipole cross section.

\begin{figure}[htbp]
    \centering
    \includegraphics[width=0.8\linewidth]{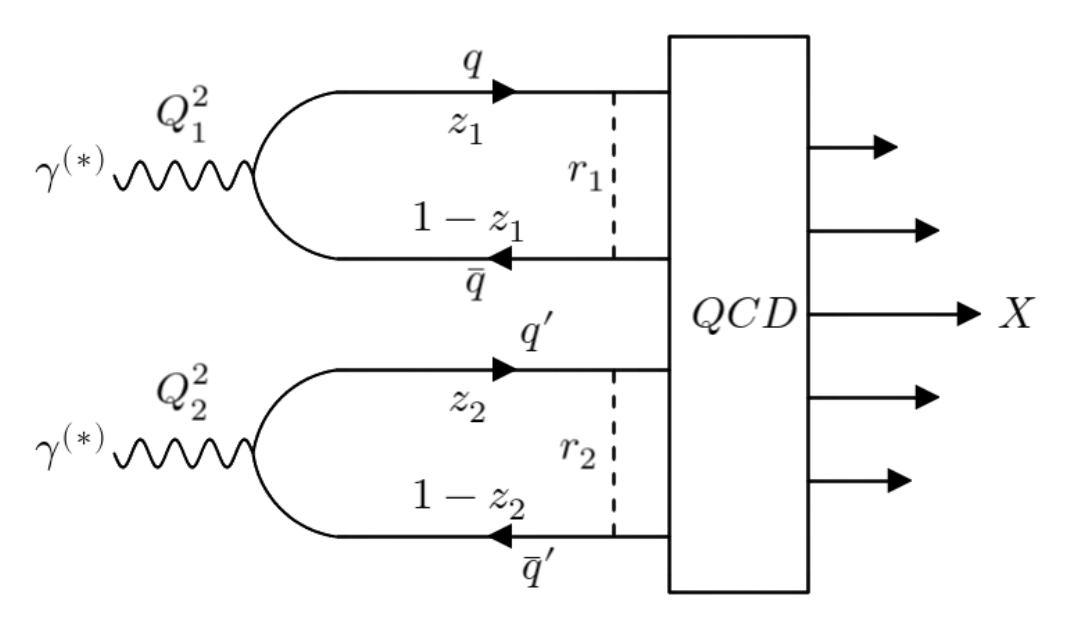}
    \caption{Pictorial diagram of the dipole-dipole cross section, where the square area represents the QCD dynamics responsible for hadron production in the final state ($X$).}
    \label{fig:dipole-dipole}
\end{figure}

The strong interaction between two photons is \linebreak illustrated in Fig.~\ref{fig:dipole-dipole}, where each photon with virtuality $Q^2_{1,2}$ fluctuates into a quark-antiquark dipole with transverse size $r_{1,2}$. The variables $z_{1,2}$ and $(1 - z_{1,2})$ represent the momentum fractions of the photons carried by the quarks and antiquarks in the dipoles, respectively.

In a full QCD treatment, the scattering of two color dipoles would be described by a hierarchy of coupled evolution equations (the Balitsky–JIMWLK hierarchy) that accounts for the mutual splitting and recombination of gluons in both objects. Solving these equations for two extended systems is analytically and numerically prohibitive. To make progress, the literature has adopted a pragmatic shortcut: the dipole–dipole cross section $\sigma_{dd}$ is replaced by an \textit{ansatz} that re‑uses the well‑known dipole–proton amplitude $\mathcal{N}(r,Y)$ (extracted from DIS). The two main prescriptions, TKM and IKT (detailed in Sec.~\ref{sec:dipole-dipole}), differ in how they map the two dipoles onto an effective dipole–target configuration. Although this reduction is not derived from QCD, it has been shown to reproduce the available $\gamma\gamma$ data from LEP after adjusting the free parameters ($\Lambda_c$ in IKT, $m_l$ in TKM).

The gluonic contribution to the $\gamma^{(\ast)}\gamma^{(\ast)}$ cross section is obtained by analogy with the $\gamma^{(\ast)}p$ case in DIS. Therefore, we can write the cross section for the interaction of two photons arising from the exchange of gluon degrees of freedom as~\citep{Timneanu:2002}:
\begin{align}\label{eq:sig_gluon}
\sigma_{gluon}^{\gamma^{(\ast)} \gamma^{(\ast)}} (W^2, Q^2_{1,2}) \! = \! & \sum_{\alpha, \beta}^{T,L} \sum_{a,b}^{N_f}\int_0^1 dz_1\int d^2{r}_1|\Psi_\alpha^a(z_1,{r}_1)|^2
\notag \\ & \times \!\! 
\int_0^1 \! dz_2\int \! d^2 {r}_2|\Psi_\beta^b(z_2,{r}_2)|^2\sigma_{a,b}^{dd}(r_1,r_2,Y).
\end{align}
{Physically, this master equation frames the high-energy interaction as a two-step quantum process: each photon first fluctuates into a color-singlet quark-anti-quark dipole with transverse size $r_i$ and momentum fraction $z_i$, weighted by the absolute square of its light-cone wave function $|\Psi|^2$, and these extended structures subsequently scatter off one another via the universal dipole-dipole cross section $\sigma^{dd}_{ab}$, which encapsulates the dense non-linear gluon dynamics. }

{It is important to emphasize that substituting the dipole-proton amplitude $\mathcal{N}(r,Y)$ directly into the dipole-dipole cross section constitutes a pragmatic phenomenological reduction rather than a first-principles derivation. By employing this mapping, the formalism implicitly assumes that a color-singlet quark-anti-quark pair interacting with another identical color-singlet state behaves identically to its interaction with a baryon target containing three valence quarks and a dense sea of multi-parton correlations. While this assumption lacks formal justification from QCD field equations, it serves as an indispensable baseline; isolating where this mapping holds—and where it breaks down under the mutual fluctuations of two gauge-boson fields—provides the exact diagnostic tool required to test the limits of universality.}

The $\sigma_{a,b}^{dd}$ is the dipole-dipole cross section, which carries all the QCD dynamics, the sum is performed over the photon polarization states $(\alpha, \beta = T, L)$ and over the quark flavor indices $(a, b)$ correspond to $u, d, s, c, b$, with this being the first time that heavy quarks are included in such an analysis in the literature. This inclusion is particularly relevant for future high-energy colliders, especially at small virtuality and high energies, where obtaining signals from Higgs boson decays and potentially new physics requires an accurate estimation heavy flavor production background via two-photon processes \citep{Boyko:2023}. Although the coherence length of the heavy dipole fluctuation is smaller, $l_\text{coh} = 2/(M_p x)$, it still needs to be much longer than the nuclear diameter (i.e. $l_\text{coh} \gg 2R_p$). This condition is only satisfied when the Bjorken-$x$ of the heavy flavor fulfills $x \ll 1/(M_p R_p)$. The results presented in this paper demonstrate that, in the energy regime of future colliders, the effects of heavy dipoles become relevant for the production of the final state. 

The Eq.(\ref{eq:sig_gluon}) involves integrations over the momentum fractions $z_{1,2}$ and dipole transverse sizes $r_{1,2}$. The photon wave function $|\Psi(z, r)|^2$ represents the probability density for the photon to fluctuate into a dipole of size $r$. The dipole-dipole cross section encodes the full QCD dynamics of the interaction and depends on the rapidity of the process, $Y = \ln(1/x)$.

In light-cone coordinates the Feynman rules are applied to calculate the light-cone wave functions, illustrated in Fig.(\ref{fig:dipole-dipole}). The wave functions corresponding to transversely (T) and longitudinally (L) photon polarized state, were first derived in perturbative quantum electrodynamics (pQED) by \citep{Nikolaev:1991, Mueller:1990}, and are given by:
\begin{subequations} \label{eq.photon_wave_function}
\begin{align}
|\Psi_T^{\gamma \rightarrow q\bar{q}}(r,z;Q^2)|^2  = & \frac{2N_c \alpha_{em}}{4\pi^2} \sum_q e^2_q \biggl[  m^2_q \, K_0^2(\bar{Q}_qr) 
  \notag \\  &  
    + [z^2 + (1-z)^2]\bar{Q}^2_q \, K^2_1(\bar{Q}_qr) \biggr] \ ,  
\\ 
|\Psi_L^{\gamma \rightarrow q\bar{q}}(r,z;Q^2)|^2 = &  \frac{8 N_c \alpha_{em}}{4\pi^2} \sum_q e^2_q
\notag \\  & \times
Q^2z^2(1-z)^2 \, K_0^2(\bar{Q}_qr) \ ,
\end{align}
\end{subequations}
where $\bar{Q}_q^2 = z(1 - z)Q^2 + m_q^2$ includes the quark mass $m_q$, $K_\nu(x)$ is the modified Bessel function of the second kind of order $\nu$, $N_c = 3$ is the number of colors, and $\alpha_{\text{em}} = 1/137$ is the electromagnetic coupling constant. {Mathematically, the modified Bessel functions $K_0$ and $K_1$ control the exponential transverse confinement of the dipole field, ensuring that fluctuations with large sizes or heavy quark masses are dynamically suppressed unless compensated by high virtuality or energy.}

The two-photon cross section with the gluon contribution requires a threshold correction factor in Eq.~(\ref{eq:sig_gluon}) in order to extend the validity of the gluon term into the larger-$x$ region.

\begin{equation}\label{eq.sig_G}
\sigma_{G}^{\gamma^{(\ast)} \gamma^{(\ast)}}(W^2, Q_{1,2}^2) = \sigma_{gluon}^{\gamma^{(\ast)} \gamma^{(\ast)}}(W^2, Q_{1,2}^2)(1-\bar{x})^{\eta_c} \ ,
\end{equation}
with $\eta_c = 2N_f - 1$, where $N_f$ is the number of active quark flavors, and where $\bar{x}$ depends on the specific dipole-dipole cross section model used. In the TKM and IKT models, the modified Bjorken-$x$ is defined differently to reflect their respective treatments of kinematics and energy dependence. {Physically, this empirical $(1-\bar{x})^{\eta_c}$ phase-space damping factor acts as an essential safeguard, smoothly turning off the gluonic dipole exchange as Bjorken-$x$ approaches unity and preventing unphysical contributions in the elastic threshold limit where multi-parton cascading ceases to operate.} The total photon-photon cross section is then given by:
\begin{equation}\label{eq.Sig_total}
\begin{aligned}
\sigma_{total,ij}^{\gamma^{(\ast)} \gamma^{(\ast)}}(W^2, Q_{1,2}^2)
&= \sigma_{G,ij}^{\gamma^{(\ast)} \gamma^{(\ast)}}(W^2, Q_{1,2}^2) \\
&\quad + \sigma_{R}^{\gamma^{(\ast)} \gamma^{(\ast)}}(W^2, Q_{1,2}^2)\,\delta_{iT}\delta_{jT} \\
&\quad + \sigma_{QPM,ij}^{\gamma^{(\ast)} \gamma^{(\ast)}}(W^2, Q_{1,2}^2).
\end{aligned}
\end{equation}
where it includes the complete contributions from each polarized state, $i,j = T,L$, with the cross section that dominates hadron production at higher energies arising from the gluonic exchange component, $\sigma_G$, which is given in Eq.(\ref{eq.sig_G}). Also the contribution at low energies ($6 \text{ GeV} \lesssim W_{\gamma \gamma} \lesssim 50 \text{ GeV}$), the interactions are dominated by the VDM (Reggeon exchange) cross section, $\sigma_{R}$, and by the Quark Parton Model cross section, $\sigma_{QPM}$, which are expressed in Eq.(\ref{eq.Sig_reg}) and Eq.~(\ref{eq.QPM}), respectively. {Conceptually, Eq.~(\ref{eq.Sig_total}) unifies the three pillars of high-energy photon-photon physics—soft non-perturbative Reggeon background, point-like quark-parton box transitions, and non-linear gluon saturation—providing a complete spectrum from threshold to asymptotic regimes.}

The two prescriptions for the dipole–dipole cross section, TKM and IKT, which will be detailed in Sec.~\ref{sec:dipole-dipole}, represent the core phenomenological ambiguity of this framework. Their systematic comparison, performed in the companion paper~\citep{ZardoBecker2026b}, will quantify the spread of predictions and define the quantitative benchmark for future experiments.

\section{Dipole-dipole cross section}\label{sec:dipole-dipole}

As noted in Sec.~2.3, the gluon contribution is expressed in terms of the dipole–dipole cross section $\sigma_{dd}$. Because a first‑principles QCD treatment of two interacting dipoles is not available, the literature has developed two phenomenological prescriptions – TKM and IKT – that relate $\sigma_{dd}$ to the dipole–proton cross section $\sigma_{dp} = 2\pi R_p^2 \mathcal{N}(r,Y)$. 

Throughout this work, we adopt the following operational definitions: a dipole is considered `small' when its transverse size satisfies $r \ll 1/Q_s(Y)$, where $Q_s(Y)$ is the saturation scale; it is considered `large' when $r \gg 1/Q_s(Y)$. The transition region $r \sim 1/Q_s(Y)$ corresponds to the saturation boundary, where the scattering amplitude $\mathcal{N}(r,Y)$ interpolates between the dilute (color transparency) and dense (saturation) regimes. {Operationally, this means that while small dipoles penetrate targets without hindrance due to color transparency, dipoles reaching or exceeding the inverse saturation scale encounter a dense wall of gluons where non-linear recombination taming the cross-section growth becomes dominant.} These definitions are model-dependent through $Q_s(Y)$, but they provide a consistent criterion for comparing TKM and IKT across the five saturation models used in this work.

In this work, we define `faster growth' quantitatively as the logarithmic derivative of the cross section with respect to the c.m. energy: $\Delta(W) \equiv d\ln\sigma/d\ln W$. {Physically, this derivative measures the effective intercept of the interaction, dictating how rapidly the cross-section rises as higher gluon densities are probed at larger center-of-mass energies.} The TKM prescription systematically yields a larger value of $\Delta(W)$ than IKT for all five saturation models tested, with the difference increasing with $W$ in a monotonic fashion. The precise numerical values of $\Delta(W)$ for each model and prescription is presented in the companion paper~\citep{ZardoBecker2026b}, where we compare our numerical results with available LEP data and with projections for future colliders.

In the TKM prescription, this relation is built around an effective dipole size $r_{\text{eff}} = r_1 r_2 / \sqrt{r_1^2 + r_2^2}$, which symmetrically combines the two dipole sizes. In the IKT prescription, the smaller dipole is treated as a projectile that scatters off the larger dipole, approximated as a static target of area $\pi r_{\text{max}}^2$. Both prescriptions neglect impact‑parameter dependence – an approximation justified for total cross sections but potentially insufficient for exclusive processes. Despite their heuristic nature, these prescriptions have been successfully tested against LEP data after adjusting the free parameters ($\Lambda_c$ for IKT, $m_l$ for TKM). {A critical examination of the IKT prescription reveals that its foundational asymmetry—forcing an arbitrary designation of projectile and target through a step function and a momentum cutoff—is not merely a formal artifact, but a reflection of a deep kinematic hierarchy. When the colliding photons possess vastly disparate virtualities, such that $Q_1^2 \gg Q_2^2$, the harder photon fluctuates into a highly compact, point-like dipole that probes the extended color field of the softer, larger configuration without inducing reciprocal saturation. However, when the virtuality become comparable, particularly in the real-photon limit ($Q_1^2 \approx Q_2^2 \approx 0$), this asymmetric imposition fails to reflect the mutual, symmetric exchange of field quanta mandated by field-theoretic crossing symmetry. Recognizing IKT as a physically motivated virtuality-ordered approximation rather than a universal description allows us to isolate precisely where asymmetric parameterizations cease to be valid and where a fully symmetric two-dipole kernel becomes mandatory.}

A subtle but important point is that both prescriptions reuse the same dipole–proton amplitude $\mathcal{N}(r,Y)$ that was extracted from DIS on protons. By doing so, they implicitly assume that a color‑singlet $q\bar{q}$ pair (a dipole) scatters off another dipole exactly as it would off a proton – an assumption that lacks theoretical justification. Furthermore, the mapping in TKM (effective size) and IKT (smaller dipole as projectile) explicitly breaks the $1 \leftrightarrow 2$ exchange symmetry when the two dipoles have comparable sizes: the cross section depends on which dipole is arbitrarily designated as the target and which as the projectile. {In the fundamental formulation of QCD, a symmetric treatment requires the interaction to be invariant under the interchange of projectile and target indices; the artificial breaking in these models reflects our current reliance on mapping a single-target DIS framework onto a symmetric two-photon collision.} In a genuine QCD treatment, the evolution of both dipoles in rapidity would be symmetric, and the scattering amplitude would depend on the sum of their individual rapidities, $Y_1+Y_2$, rather than on an arbitrary assignment. The present work accepts these limitations as part of the state of the art; the systematic uncertainty they introduce is quantified in the companion paper~\citep{ZardoBecker2026b}.

\subsection{A possible interpretation of the broken symmetry}
The asymmetry in the IKT prescription (the cross section depends on which dipole is designated as the projectile) is often viewed as a formal defect. However, one may reinterpret it as a natural consequence of the fact that the two photons can have very different virtuality $Q_1^2$ and $Q_2^2$. In the limit $Q_1^2 \gg Q_2^2$, the first photon fluctuates into a much smaller dipole ($r_1 \sim 1/Q_1$) than the second ($r_2 \sim 1/m_q$). In this regime, the physical picture of a small projectile scattering off a large target is justified, and the IKT prescription reduces to $\sigma_{dd} \approx 2\pi r_2^2 \mathcal{N}(r_1)$. {Here, the large dipole acts as a static color field source of geometric cross-sectional area $\pi r_2^2$, through which the compact, point-like probe $\mathcal{N}(r_1)$ navigates without suffering internal saturation corrections.} Conversely, when $Q_1^2 \sim Q_2^2$, both dipoles are comparable in size, and the asymmetry becomes problematic. Thus, the IKT prescription may be viewed as a \textit{virtuality‑ordered approximation}: it is expected to work best when one photon is much harder than the other, while TKM, being symmetric, is preferable when the two virtuality are comparable. A direct test of this hypothesis would be to compare predictions for $\sigma^{\gamma^*\gamma^*}(W,Q^2)$ with $Q_1^2 = 10Q_2^2$ – a case where the two prescriptions should give very different results, and the data could reveal which ordering is physically preferred.

It is worth noting that the IKT prescription, despite its broken exchange symmetry, has been shown to describe $\gamma p$ data successfully in the regime where the photon virtuality is large and the proton provides a large target \citep{Basso:2008}. This suggests that the asymmetry in IKT may be physically motivated in processes where one of the interacting systems is much smaller than the other – a point often overlooked in critical assessments of the prescription.

The quantitative comparison between the two prescriptions, including a detailed study of their phenomenological consequences and a proposal for experimental discriminants, is presented in the companion paper~\citep{ZardoBecker2026b}.

\subsection{TKM Prescription}

The TKM prescription postulates that the interaction between two dipoles can be mapped onto the scattering of a single effective dipole of size $r_{\text{eff}}$ off a hadron target, using the same amplitude $\mathcal{N}$ as in DIS. Concretely, the dipole–dipole cross section written by Tîmneanu–Kwieciński–Motyka \citep{Timneanu:2002} is formulated using an effective dipole size and a modified Bjorken-$x$ variable:
\begin{align}\label{eq.sdd_TKM}
&\sigma^{dd}_{a,b}(r_{1,2}, \bar{x}_{ab}) = \sigma^{a,b}_0 \mathcal{N}(r_{eff},\bar{x}_{ab}) \ ,
\end{align}
where $\bar{x}_{ab}$ is a symmetric expression of the combination of quark flavors and virtuality of both photons
\begin{align}
     &\bar{x}_{ab} = \frac{Q^2_1 + Q^2_1 + 4m_a^2 + 4m_b^2}{W^2 + Q^2_1 + Q^2_2} \ ,
\end{align}

This expression allows the model to be extended down to the limit $Q^2_{1,2} \approx 0$. The mass of the light quarks, $m_l$, is treated as a free parameter in the model and is adjusted to achieve the best agreement between the predicted photon--photon cross section and the experimental data available in the energy range $6~\text{GeV} \leq W_{\gamma\gamma} < 160~\text{GeV}$~\citep{Goncalves:2012}.

The effective dipole size, $r_{\text{eff}}$, is defined in terms of the sizes of the two interacting dipoles as:
\begin{equation}
r^2_{eff} = \frac{r^2_1 r^2_2}{r^2_1 + r^2_2} \quad \text{and} \quad \sigma^{a,b}_0 = \frac{2}{3}\sigma_0 \ ,
\end{equation}
{Mathematically, this harmonic-mean combination ensures that $r_{\text{eff}}$ is always dominated by the smaller of the two dipoles when sizes differ significantly, while scaling symmetrically when both dipoles approach equal dimensions.} Thus, small or large values of $r_{\text{eff}}^2$ indicate that both dipoles are correspondingly small or large. In contrast, intermediate values of $r_{\text{eff}}^2$ correspond to the interaction between a mixing of a small dipole and a large one. The factor $\sigma^{a,b}_0$ in the dipole–dipole cross section is related to the constant $\sigma_0 = 2\pi R_p^2$ that appears in the dipole–proton scattering amplitude models. According to~\citep{Timneanu:2002}, the ratio between these constants should reflect the difference in the number of valence constituents of the target: a dipole contains two valence quarks, while the proton contains three.

\subsection{IKT Prescription}

The IKT prescription, originally proposed for Mueller–Navelet~\citep{Mueller:1990} jets, assumes that the smaller dipole acts as a projectile and the larger one as a static target. The dipole–dipole cross section is then approximated by the geometric area of the larger dipole times the dipole–proton amplitude of the smaller one, weighted by a step function. This yields:

\begin{align}\label{eq:sigmadd-def}
\sigma_{a,b}^{dd}(r_1,r_2,\bar{Y}) &= 2\pi r_1^2 \mathcal{N}(r_2,\bar{Y}_2)\Theta(r_1-r_2) 
\notag  \\ & 
+ 2\pi r_2^2 \mathcal{N}(r_1,\bar{Y}_1)\Theta(r_2-r_1) \ , 
\end{align}
\begin{align}
\sigma_{a,b}^{dd}(r_1,r_2,\bar{Y}) & =  2\pi r_1^2  \mathcal{N}(r_2,\bar{Y}_2)\Theta(r_1-r_2) \notag \\ & \quad  + 2\pi r_2^2 \mathcal{N}(r_1,\bar{Y}_1)\Theta(r_2-r_1),
\end{align}
the pseudo-rapidity $\bar{Y}_i^f=\ln(1/\bar{x}_i^f)$ is written with Bjorken $x$ in the form:
\begin{equation}
\bar{x}_i^f=\frac{Q_i^2 + 4m_f^2}{W^2	+ Q_i^2} \ .
\end{equation}

\begin{figure}[htbp]
    \centering
    \includegraphics[width=1.0\linewidth]{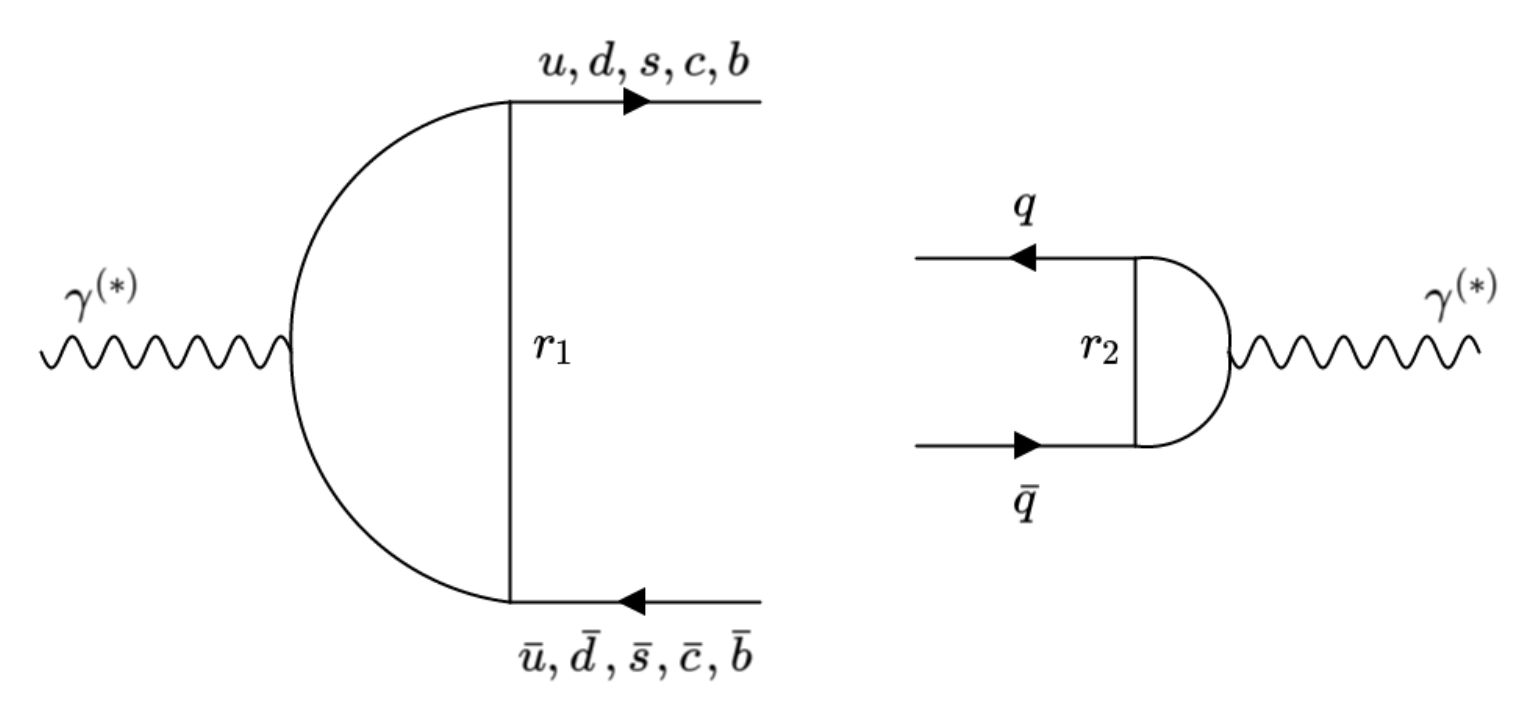}
    \caption{Pictorial diagram of the IKT prescription for the case $r_1 > r_2$, where the bigger dipole acts as the target and includes all quark flavors. The smaller dipole, considered as the projectile, has its flavor content associated with the variable $\bar{Y}$ used in the evolution of the dipole scattering amplitude.}\label{fig.DD_heaviside}
\end{figure}

The IKT prescription considers the smallest dipole as the projectile, which carries the scattering amplitude, and the largest dipole as the target, with a cross-sectional area of $\pi r_{1,2}^2$. For example, in the case where $r_1 > r_2$, the dipole-dipole interaction can be illustrated in Fig.(\ref{fig.DD_heaviside}). 

According to~\citep{Goncalves:2012}, due to the quadratic dependence on the target dipole size in Eq.(\ref{eq:sigmadd-def}), contributions from large values of $r_1$ and $r_2$ become significant in the cross section. To ensure that the interaction remains within the perturbative QCD regime, it is necessary to implement an upper cutoff ($r_{\text{max}}$) on the dipole sizes in the integration over $r_{1,2}$. As proposed in~\citep{Goncalves:2007}, the maximum dipole size is expected to be inversely proportional to the QCD energy scale, i.e., $r_{\text{max}} = 1/\Lambda \approx 1/\Lambda_{\text{QCD}}$. Here, $\Lambda$ is understood as a sharp boundary between perturbative and non-perturbative physics and is the only free parameter in this prescription, while all other parameters associated with the dipole–proton scattering amplitude are kept fixed. {Physically, this cutoff prevents unphysical contributions from extremely large, soft configurations where perturbation theory and color transparency assumptions break down.}

Varying the momentum cutoff $\Lambda_c$ within a standard $\Lambda_{\text{QCD}}$ range demonstrates that the cross-section hierarchy $\sigma_{\text{TKM}} > \sigma_{\text{IKT}}$ remains robust across all tested values. A comprehensive numerical quantification of this sensitivity and its impact on the TKM--IKT spread is deferred to the companion paper~\citep{ZardoBecker2026b}.

If the value of $\Lambda$ required to fit the two-photon real cross section data were much smaller than $\Lambda_{\text{QCD}}$, this would indicate that non-perturbative effects should be included in the prescription. With the parameter values obtained in this analysis, however, the need to implement non-perturbative corrections was not identified. Nevertheless, some authors suggest introducing a correction term in the photon wave function to account for such effects, which are attributed to the large transverse size of dipoles resulting from quasi-real photon fluctuations~\citep{Goncalves:2020,Shi:2023}.

\section{Non-linear evolution equation}\label{sec:BK_models}

In the present work, we analyze the dipole--dipole cross section using five different models for the dipole--proton scattering amplitude. These models are based either on phenomenological fits or on the asymptotic solutions of the BK equation, which is an integro-differential, non-linear evolution equation that incorporates saturation effects at high energies is written as ~\citep{Balitsky:1996,Kovchegov:1999}

\begin{align}\label{eq.BK}
& \frac{\partial\mathcal{N}(r,x)}{\partial\log (x_0/x)} = \int d \mathbf{r_1} K(\mathbf{r},\mathbf{r}_1,\mathbf{r}_2)
\notag \\ & \quad
\times \biggl[ \mathcal{N}(r_1,x) + \mathcal{N}(r_2,x)
- \mathcal{N}(r,x) - \mathcal{N}(r_1,x)\mathcal{N}(r_2,x)\biggr] \ ,
\end{align}
where ${r}_2 = {r} - {r}_1$, $x_0$ is the Bjorken value of $x$ from the initial condition where the evolution begins and the kernel $K(\mathbf{r},\mathbf{r}_1,\mathbf{r}_2) = \mathbf{r}^2/\mathbf{r}_1^2 \mathbf{r}_2^2$  is related to the differential probability amplitude
of a dipole $\mathbf{r}$ splitting into two dipoles $\mathbf{r}_1$ and $\mathbf{r}_2$.
To evaluate the phenomenological spread of this saturation dynamics, we test five distinct models for the initial condition and non-linear evolution, establishing an essential benchmark that we quantify in the companion paper.

The five models considered in this contribution are not meant to represent a unique theoretical description; rather, they span the current theoretical envelope—each embodying different assumptions about the saturation scale, the running coupling, and the transition between dilute and dense regimes. {The evolution of the dipole scattering amplitude $\mathcal{N}(r,Y)$ across this landscape is governed by non-linear equations, such as the Balitsky-Kovchegov (BK) equation, which resumes leading logarithms in energy. Although no explicit time coordinate appears in the light-cone formalism, evolution proceeds implicitly through the rapidity $Y = \ln(1/x)$ (or equivalently $Y = \ln(W^2 / Q^2)$). Physically, this rapidity acts as an internal quantum-mechanical ``evolution time,'' dictating how the phase space of soft gluon emissions expands as the center-of-mass energy increases. As the number of partons multiplies into large values of $Y$, the probability densities for gluon emission and gluon recombination eventually become comparable; these competing non-linear recombination effects trigger saturation, taming the unphysical growth of the cross section and driving the transition from a dilute parton gas to a dense gluon condensate.}

The scattering amplitude $\mathcal{N}$ is related to the $S$-matrix through the relation $S = 1 - \mathcal{N}$, with the unitarity condition requiring that $\mathcal{N} \leq 1$. This constraint is naturally satisfied by the solution of the BK equation. The regime $\mathcal{N} \ll 1$ corresponds to the dilute limit and is separated from the saturation regime by the saturation scale $Q_s^2$.  However, the dipole--proton cross section can continue to grow with energy even after the black-disk limit is reached ($\mathcal{N} = 1$), indicating that unitarity corrections must be carefully accounted for, particularly in the high-density regime.

\begin{figure}[htbp]
    \centering
    \includegraphics[width=1.0\linewidth]{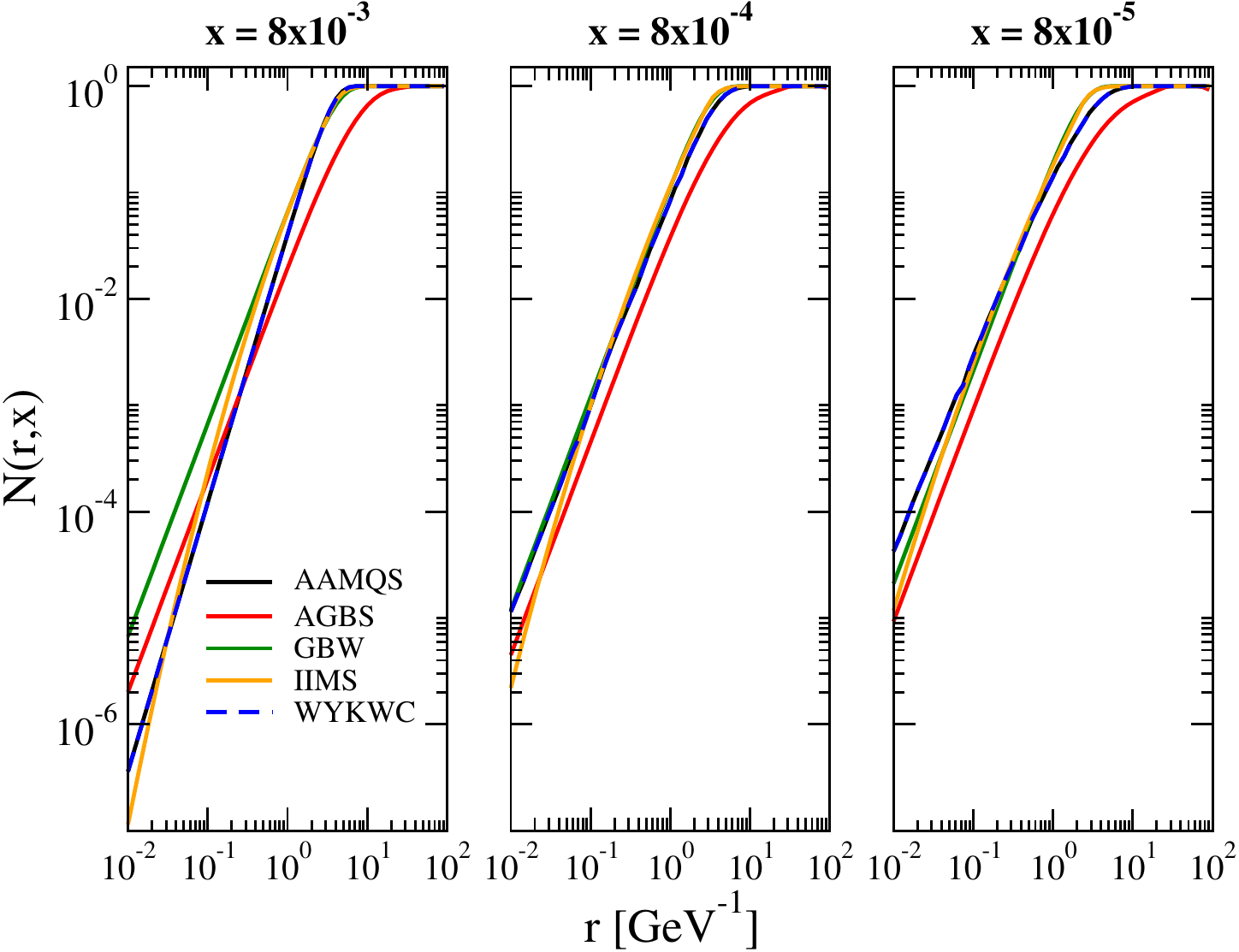}
    \caption{The dipole-proton scattering amplitude as a function of dipole size is shown for different values of Bjorken-$x$: $x = 8 \times 10^{-3}$ (left), $x = 8 \times 10^{-4}$ (center), and $x = 8 \times 10^{-5}$ (right). The plots highlight that, as the energy increases (i.e., $x$ decreases), smaller dipoles become increasingly relevant in the scattering amplitude. The models are more distinguishable at larger values of $x$ and smaller dipole sizes $r$. Most models tend to saturate in the same region of dipole size, except for AGBS, which exhibits a lower amplitude and saturates at a larger value of $r$.}\label{fig:N_coordenate_space}
\end{figure}

Thus, it is important to perform further tests on the models to probe their universality. All dipole--proton scattering amplitudes used in this work are shown in Fig.~(\ref{fig:N_coordenate_space}). These include those formulated in coordinate space~\citep{Albacete:2011, GolecBiernat:1998, Soyez:2006, Soyez:2007}, as well as those formulated in momentum space~\citep{Amaral:2021}.

Based on the behavior of the dipole--proton scattering amplitude, which exhibits both color transparency and saturation, it is possible to gain insight into the two-photon cross section, as expressed in Eq.~(\ref{eq:sig_gluon}). As is well known, for small dipole sizes, $r \ll 1/Q_s(Y)$, the cross section vanishes as $\sigma \sim r^2$, reflecting the color transparency regime described by the GBW model. In contrast, for large dipoles, $r \gg 1/Q_s(Y)$, the cross section saturates to a constant value. This behavior is clearly illustrated in Fig.~\ref{fig:N_coordenate_space}.

The gluonic contribution to $\gamma\gamma$ interactions is typically modeled in analogy with $\gamma p$ interactions. However, the cross section for real photons, $\sigma^{\gamma\gamma}(W^2) \sim \ln^2(W^2/W_0^2)$, grows more rapidly with energy than the corresponding $\gamma p$ cross section, $\sigma^{\gamma p}(W^2) \sim \ln(W^2/W_0^2)$. This steeper energy dependence highlights the even more significant role of QCD dynamics in two-photon interactions at high energies.

In the case of virtual photons, the cross section behaves as $\sigma^{\gamma^\ast \gamma^\ast}(W^2, Q^2) \sim \ln^2(Q_s^2/Q^2)$, exhibiting a double-logarithmic enhancement relative to $\sigma^{\gamma^\ast p}(W^2, Q^2) \sim \ln(Q_s^2/Q^2)$. The quantity $\ln^2(r^2 Q_s^2)$ measures how far the probing dipole scale is from the saturation regime. If we set $r = 1/Q$, then $Q \sim Q_s$ corresponds to the transition region where the logarithm vanishes. For $Q \ll Q_s$, the system lies deep within the saturation regime, while for $Q \gg Q_s$, the system is in the dilute regime where the gluon density grows rapidly with energy. This transition is ensured through the property of geometric scaling ~\citep{Iancu:2008}, a property originally observed in the total cross section of Deep Inelastic Scattering (DIS) \citep{Stasto:2001} . Geometric scaling refers to the fact that observables depend predominantly on the scaling variable $\tau = Q/Q_s(Y) = rQ_s(Y)$, which combines the $Q$ and $Y$ into a single variable, rather than depending on them separately.

In the asymptotic limit $W \to \infty$ (equivalently $Y \to \infty$), both TKM and IKT prescriptions tend to the black-disk limit where $\mathcal{N}(r,Y) \to 1$ for all dipoles with $r \gtrsim 1/Q_s$. However, the difference between the two prescriptions does not necessarily vanish in this limit: the IKT cross section retains a factor $2\pi r_{\text{max}}^2$ from the larger dipole, which is cut off by $\Lambda_c$. Thus, even at asymptotically high energies, the TKM–IKT ambiguity persists, albeit with a reduced magnitude. This asymptotic behavior will be explored qualitatively in the companion paper.

{\subsection{Comparison of models in momentum space}\label{sec:compare_N(k)*}}

\begin{figure}[htbp]
    \centering
    \includegraphics[width=1.0\linewidth]{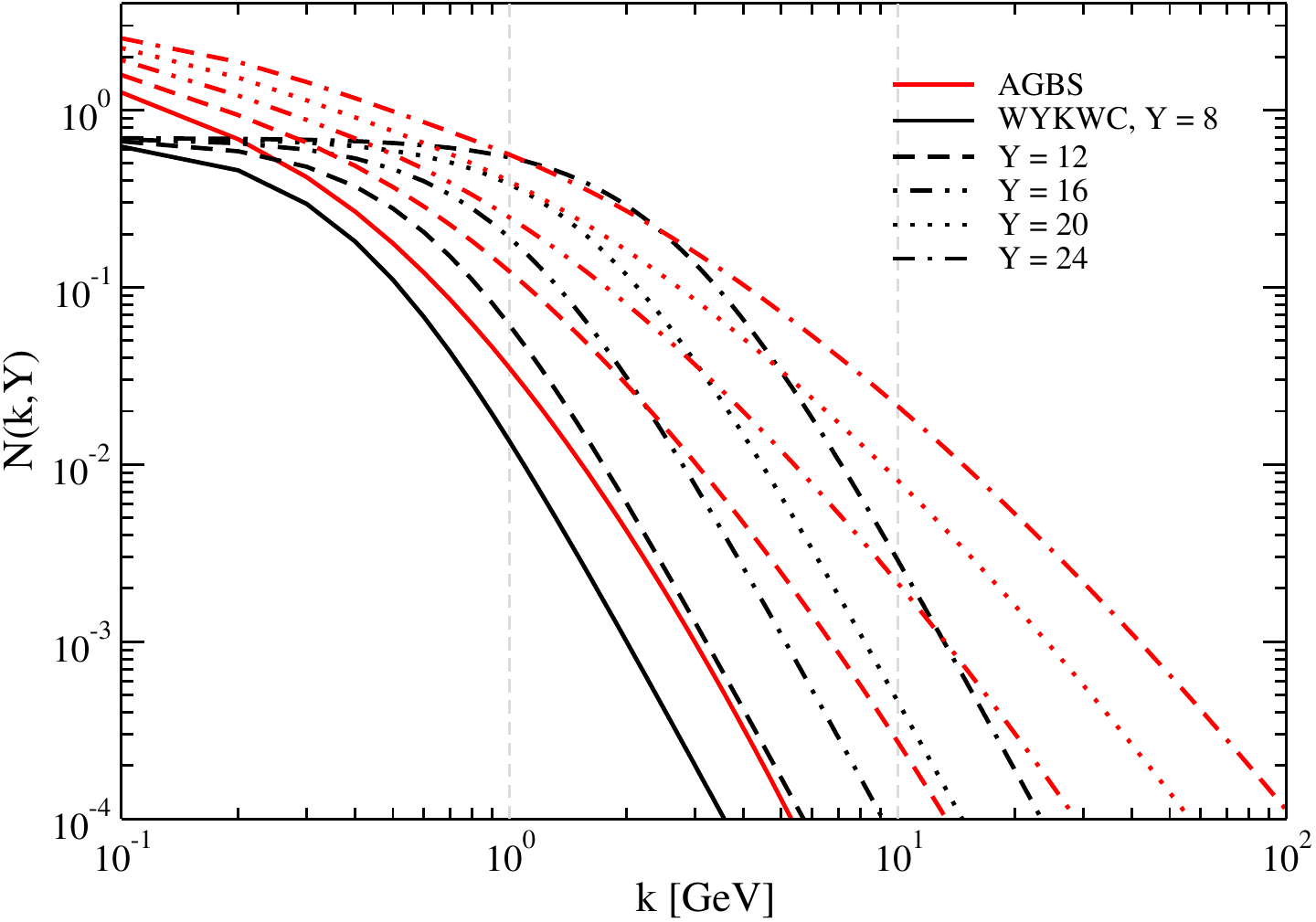}
    \caption{{The dipole-proton scattering amplitude in momentum space, $\tilde{\mathcal{N}}(k)$, is shown for the AGBS and WYKWC models at different values of rapidity. Non-linear effects due to gluon overlap become evident when $\mathcal{N}(k) \approx 1$, which occurs at low transverse momentum $k$. In contrast, the amplitude decreases and the system enters the dilute regime, when the wavefront extends to larger $k$.
}}
    \label{fig:N_momentum}
\end{figure}

{The behavior of the dipole scattering amplitude in momentum space for the AGBS and WYKWC models (which are detailed in the next section) are shown in the Fig.(\ref{fig:N_momentum}). The main difference between them is that the WYKWC model exhibits explicit saturation effects at low values of $k$, while the AGBS model describes a higher parton density and extends the dilute regime to higher transverse momenta. }
{A subtle yet critical mathematical challenge in unifying coordinate-space saturation models with momentum-space asymptotic solutions lies in the integrity of the mapping between representations. Because observables are evaluated in coordinate space while certain advanced BK solutions are naturally governed in momentum space, the application of a modified Hankel transform involving the Bessel function of the first kind introduces non-trivial dependencies on the infrared and ultraviolet boundaries. To ensure that this transition does not violate the fundamental unitarity bound, $\mathcal{N} \le 1$, or introduce artificial normalization drifts, the effective momentum scale $k_{\text{eff}}$ in the symmetric TKM prescription must be established via stationary-phase evaluations of the transform integral. This ensures that the geometric scaling properties and wavefront propagation derived in momentum space are mapped onto coordinate-space configurations without distorting the underlying gluon distribution, thereby securing a mathematically self-consistent bridge across dual-representation saturation frameworks.}

{In this work, all observables are formulated in coordinate space. To relate them to momentum space, we use a modified Fourier transform, Eq.(\ref{eq:Tr-fourier}). In the TKM prescription, we introducing the effective dipole size $r_{\text{eff}}$ and define an effective momentum scale $k_{\text{eff}}$.}

{To ensure that momentum-space amplitudes, such as those derived from the AGBS and WYKWC models, are consistently incorporated into the coordinate-space dipole cross section without normalization ambiguities, we apply the modified Fourier transform relating $\mathcal{N}(r,Y)$ and \(\tilde{\mathcal{N}}(k,Y)\) as defined in Eq.~(\ref{eq:Tr-fourier}). In the context of the TKM prescription, where an effective dipole size $r_{\text{eff}}$ is utilized, the corresponding effective momentum scale $k_{\text{eff}}$ is established through the stationary-phase evaluation of the Hankel transform. This mapping preserves the unitarity bound $\mathcal{N} \leq 1$ across both representations and guarantees that the normalization of the dipole cross section remains invariant under the space-to-momentum transition, thereby providing a consistent bridge between coordinate- and momentum-formulated saturation models.}

{The differences between the models—particularly in the behavior of the amplitude for small and large dipoles—will be reflected in the phenomenological predictions presented in the companion paper, where their impact on observables such as 
$\sigma^{\gamma \gamma}$ and $F_2^{\gamma}$
  is quantified.}

\vspace{0.1in}

\subsection{Coordinate space}

In coordinate space, we select three of the most well-established models for the dipole scattering amplitude: GBW, IIMS, and the numerical solution of Eq.~(\ref{eq.BK}) with running coupling, known as rcBK. These models are discussed in detail in the following section.

\subsubsection{GBW Model}
The Golec-Biernat--Wüsthoff model \citep{GolecBiernat:2018} was the simplest phenomenological model that includes saturation physics. Described in the form:
\begin{equation}\label{eq.GBW}
\mathcal{N} =  1 - \exp \left( - \frac{r^2 Q^2_s(x)}{4} \right) \ ,
\end{equation}
here, $r$ denotes the transverse dipole size, and $Q_s^2(x)$ is the $x$-dependent saturation scale, defined as \linebreak $Q^2_s(x) = Q^2_0(x/x_0 )^{-\lambda}$ with $Q_0 = 1$GeV.

The Eq.(\ref{eq.GBW}) interpolates between the dilute and dense partonic regimes. For small dipole sizes ($r \ll 1/Q_s$), the amplitude behaves as $\mathcal{N} \approx r^2$, exhibiting color transparency. In contrast, for large dipole sizes ($r \gg 1/Q_s$), the amplitude saturates and approaches a constant value, behavior how homogeneous disk: $\mathcal{N} = 1$. This saturation behavior is essential for maintaining unitarity at high energies.

A crucial element in our analysis is the assumption that the saturation scale $Q_s^2$ depends on the Bjorken-$x$ variable in such a way that, as $x$ decreases, one must probe smaller distances (i.e., higher $Q^2$) to resolve the increasingly dense parton structure of the proton.

The updated GBW parameters \citep{GolecBiernat:2018}, fitted to proton structure function from HERA, ZEUS and H1 data, are fixed as: $\sigma_0 = 27.43~\text{mb}$, $\lambda = 0.248$, and \linebreak $x_0 = 0.40 \times 10^{-4}$. The quark masses used are: \linebreak $m_{u,d,s} = 0.14~\text{GeV}$, $m_c = 1.4~\text{GeV}$ and $m_b = 4.6~\text{GeV}$.

\subsubsection{IIMS Model}

A model initially proposed by Iancu--Itakura--Munier \citep{Iancu:2004} and later modified by Soyez \citep{Soyez:2007,Soyez:2006} to include the contribution of heavy quarks, the IIMS model has demonstrated excellent agreement with HERA data for the proton structure function. The inclusion of heavy quark effects significantly improved the model's accuracy.

The IIMS model was developed to smoothly interpolate between the dilute (color transparency) regime and the saturation regime of QCD at small values of $x$. It reproduces BFKL dynamics in the dilute regime ($r Q_s \leq 2$) and incorporates saturation effects for large dipole sizes ($r Q_s \geq 2$). This transition is ensured through the property of geometric scaling. The IIMS model is write as
\begin{eqnarray}
\hspace{-0.5cm} \mathcal{N}(x,\mathbf{r}) &=& \left\{
\begin{array}{cc}
 \frac{1}{2} r Q_s \, \mathcal{N}_0 \, e^{2 \left( \gamma_s + \frac{\log(2/rQ_s)}{\kappa\lambda Y} \right)} , & rQ_s \leq 2  \\
1-e^{-a \log(brQ_s)}, & rQ_s\geq 2
\end{array}
\right. \ ,
\end{eqnarray}
the constants $a$ and $b$ appearing in the saturated regime are determined by enforcing continuity of the amplitude and its derivative at the matching point $r Q_s = 2$.
\begin{subequations}
\begin{eqnarray}
a &=& -\frac{\ln(1-\mathcal{N}_0)}{\ln(2b)^2} \ , \\
b &=& \frac{1}{2}(1 - \mathcal{N}_0)\exp\left[ \frac{\ln(2)}{\kappa \lambda Y} -\gamma_s  \right] \ .
\end{eqnarray}
\end{subequations}

In the analysis of the IIMS model with the inclusion of heavy quarks~\citep{Soyez:2007}, the parameters fitted to HERA data are fixed as $\gamma_s = 0.7376$, $\lambda = 0.1632$, and $x_0 = 3.344$. The quark masses used are $m_{u,d,s} = 140~\text{MeV}$, \linebreak $m_c = 1.4~\text{GeV}$ and $m_b = 4.5~\text{GeV}$.

\subsubsection{RCBK Model}

The Balitsky--Kovchegov evolution equation incorporating running coupling corrections was formulated in~\citep{Kovchegov:2007, Balitsky:2007, Albacete:2009}, providing a consistent implementation of small-$x$ QCD dynamics. This framework was subsequently extended to include charm and bottom quark contributions within the AAMQS analysis~\citep{Albacete:2011}.

This approach is implemented through numerical solutions of the BK equation, fitted to HERA data in the high-energy regime (small $x$, with $x_0 < 0.008$). One of the key advantages of using the rcBK framework is that its evolution kernel incorporates leading-order (LO) running coupling corrections as prescribed by Balitsky~\citep{Balitsky:2007}, expressed as:
\begin{align}\label{eq.K_RCBK}
K^{run}(\mathbf{r},\mathbf{r_1},\mathbf{r_2}) &= \frac{N_c \alpha_s(r^2)}{2\pi^2} 
\Biggl[ 
\frac{r^2}{r^2_1r^2_2} + 
\frac{1}{r^2_1}\left( \frac{\alpha_s(r^2_1)}{\alpha_s(r^2_2)}-1\right) 
\notag \\ & \hspace{1.5cm}
+ 
\frac{1}{r^2_2}\left( \frac{\alpha_s(r^2_2)}{\alpha_s(r^2_1)}-1\right) 
\Biggr] \ ,
\end{align}
where it describes the probability density for a parent dipole to split into two daughter dipoles. It has been shown that the inclusion of running coupling significantly suppresses particle number fluctuations~\citep{Dumitru:2007}. Such corrections allow for a more accurate estimate of soft gluon emissions and running coupling effects in the evolution kernel.

In the numerical implementation of the rcBK equation, the core of the code includes the expression for the number of active quark flavors ($n_f=5$), which enters through the regularization of the running coupling constant $\alpha_s$. The initial condition adopted for the dipole-proton scattering amplitude is based on the GBW model, fitted with the experimental data from HERA. The quark masses used in the calculation are $m_{u,d,s} = 0.14~\text{GeV}$, $m_c = 1.27~\text{GeV}$, and $m_b = 4.2~\text{GeV}$.

\subsection{Momentum space} 
Considering the independence of the impact parameter, it is possible to obtain the BK equation in the momentum space \citep{Kovchegov:2000,Marquet:2005}, through the Fourier transform
\begin{equation}\label{eq:Tr-fourier}
\mathcal{N}(r,Y) \! = \! r^2  \! \int \frac{d^2 k}{2\pi} e^{i{k} \cdot {r}} \tilde{\mathcal{N}} (k,Y) 
\! = \! r^2 \! \int_0^\infty   dk k J_0(kr)\tilde{\mathcal{N}}(k,Y)  \ ,
\end{equation}
where $J_0$ is a Bessel function of the first kind, which appears in the modified Fourier transform that is useful when $\mathcal{N}(r,Y)$ does not depend on the angle of ${k}$. This transform is valid for a very large nucleus; however, when the target is considered as a homogeneous disk, this approximation is sufficiently satisfied. This modified Fourier transform was used to explicitly show the behavior of the dipole scattering amplitude in momentum and coordinate space, as plotted in Fig.~\ref{fig:N_coordenate_space}.

In the momentum space, the asymptotic solutions of the BK equation exhibit traveling wave behavior in the high-energy limit ($Y \to \infty$)~\citep{Munier:2004,Marquet:2005}. These solutions provide a natural explanation for the phenomenon of geometric scaling.

The BK equation at leading logarithmic (LL) order in momentum space, whose solution is the dipole scattering amplitude $\mathcal{N}(k, Y)$, is written as:
\begin{align}
    \partial_Y \mathcal{N} = \bar{\alpha}_s \chi (-\partial_Y) \mathcal{N} - \bar{\alpha}_s \mathcal{N}^2 \ ,
\end{align}
where $\bar{\alpha}_s = \alpha_s N_c / \pi$ and $L = \log(k^2 / k_0^2)$, with $k_0$ being an infrared cutoff scale. The function $\chi(\gamma) = 2 \psi(1) - \psi(\gamma) - \psi(1 - \gamma)$ is the eigenvalue of the BFKL kernel.

After an appropriate change of variables, the BK equation reduces to the Fisher–Kolmogorov–Petrovsky–Piscounov (FKPP) equation~\citep{Fisher:1937} for a function $u(\rho, t) \propto N(k, Y)$ when the kernel is approximated using the saddle-point method i.e., expanded to second order in the derivative $\partial_L$, also known as the diffusive approximation. In this case, the equation takes the form\citep{Kovchegov:2000}:
\begin{align}
    \partial_t u (\rho,t) = \partial^2_ \rho (\rho, t) + u (\rho,t) - u^2(\rho,t) \ , 
\end{align}
with $t \sim Y$ and $\rho \sim L$ corresponding to the time and space variables, respectively. The FKPP equation admits asymptotic solutions in the form of traveling waves \citep{Munier:2004}. This means that, at large rapidity (corresponding to very small Bjorken-$x$), the solution takes the form \linebreak $u(\rho, t) = u(\rho - v_c t)$, a wavefront propagating toward larger values of $\rho$ at a constant speed $v_c$, without deformation \citep{Amaral:2021}.

\subsubsection{AGBS model}\label{sec:AGBS}

The AGBS model proposes an expression that analytically interpolates the known behaviors of $\tilde{\mathcal{N}}(k,Y)$ in infrared regions, $k\ll Q_s$,  and ultraviolet, $k\gg Q_s$ , its explicit form is given by:

\begin{align}\label{eq:agbs}
\tilde{\mathcal{N}}(k,Y) &= \left[\ln\left( \tilde{\tau} + \frac{1}{\tilde{\tau} }\right) + 1 \right](1-e^{-\tilde{\mathcal{N}}_{dil}}) \ ,
\end{align}

where
\begin{align}\label{eq:Tdil} 
\tilde{\mathcal{N}}_{dil} &= \exp\left[-\gamma_c \log\left(\tilde{\tau}^2-\frac{\mathcal{L}^2 - \log^2(2)}{2\bar{\alpha}_s \chi_c''(\gamma_c)Y} \right) \right] \ ,
\end{align}
where $\mathcal{L} = \log\left[1 + \tilde{\tau}^2 \right]$ and $\tilde{\tau}=k/Q_s(Y)$ is the geometric scaling in momentum space. The saturation scale, $Q^2_s(Y)$, is relation with associated with the position of the wavefront, 
\begin{equation}
 Q_s^2(Y) \simeq  k_0^2\exp\left(v_c Y \right) \quad | \qquad v_c=\bar{\alpha}\chi'(\gamma_c) \ ,
\end{equation}
the update parameters $\gamma_c = 0.6275$, $k_0^2 = 2.6\times10^{-3}$ GeV$^2$, $v_c = 0.136$ and $\chi_c'' = 3.4$ with $R_p = 5.1$GeV$^{-1}$ \citep{Amaral:2021}. 

The logarithmic terms in the equation correspond, respectively, to the behavior of the amplitude in the infrared and ultraviolet domains (small and large momentum), used to describe the vicinity of the saturation scale. The amplitude $\tilde{\mathcal{N}}(k, Y)$ is unitarized via the eikonal factor, $(1 - e^{-\tilde{\mathcal{N}}^{\text{dil}}})$, where $\tilde{\mathcal{N}}^{\text{dil}}$ describes the transition between the dilute and saturated regimes \citep{Amaral:2007} . The term $2 \bar{\alpha}_s \chi_c''(\gamma_c) Y$ accounts for the violation of geometric scaling due to its explicit dependence on rapidity. However, this effect becomes less significant above the saturation scale, i.e., at large values of $Y$.

\subsubsection{WYKWC Model}\label{sec:WYKWC}
The Wang--Yang--Kou--Wang--Cheng model~\citep{Wang:2021} is also based on the asymptotic analytical solution of the BK equation in momentum space, which exhibits traveling wave behavior similar to that of the FKPP equation. Using the Homogeneous Balance Method, the authors derived a new expression for the dipole--proton scattering amplitude in momentum space, achieving a successful description of vector meson production in DIS~\citep{Wang:2022}. The model is written as follows:
\begin{eqnarray}
\tilde{\mathcal{N}}(\tilde{\tau}) &=& A_0 \left[ 1 + e^{-\theta} \, \tilde{\tau}^{2 (A_0/6A_2)^{1/2}} \right]^{-2} \ ,
\end{eqnarray}
where the geometric scaling ($\tilde{\tau}$) is described in the form:
\begin{eqnarray}
\tilde{\tau}^2 &=& \frac{k^2}{Q^2_s(Y)} =  \frac{k^2}{k_0^2 \exp{\left[ \left( A_1 + 5\sqrt{A_2 A_0/6} \right)\bar{Y} \right] }} \ ,
\end{eqnarray}
here, $\bar{Y} = Y \bar{\alpha}_s$, where $\bar{\alpha}_s = \alpha_s N_c/\pi = 0.191$. The parameters of the model, fitted from $F_2$ in the HERA data \citep{Wang:2022}, which is: $A_0=0.696$, $A_1=0.661$, $A_2 = 0.112$, $\theta = -0.463$ and $R_p = 5.484$ GeV$^{-1}$ with the quark mass $m_{u,d,s} = 0.14~\text{GeV}$, $m_c = 1.4~\text{GeV}$, $\alpha_s = 0.2$ and $k_0 = 0.2~\text{GeV}^2$.

 { \section{Preliminary quantitative spread and comparison with LEP data}}\label{sec:preliminary_spread}

{To substantiate the theoretical benchmark defined by the TKM and IKT prescriptions, it is essential to examine their preliminary quantitative behavior against available experimental data from LEP. When evaluated using representative saturation models—such as the coordinate-space rcBK and GBW frameworks alongside momentum-space formulations like AGBS—the TKM and IKT prescriptions generate a distinct theoretical band for the total photon-photon cross section $\sigma_{\gamma\gamma}(W)$. In the low-to-intermediate energy regime spanning $W \lesssim 160$ GeV, which corresponds to the region covered by LEP measurements, both TKM and IKT predictions remain constrained within a relatively narrow envelope, successfully describing the magnitude and general trend of the data when combined with the VDM and QPM components. As the center-of-mass energy increases toward the TeV scale, however, the two prescriptions begin to diverge systematically. This divergence stems directly from their distinct underlying mappings, contrasting TKM's symmetric effective size $r_{\text{eff}}$ with IKT's asymmetric projectile-target hierarchy regulated by the momentum cutoff $\Lambda_c$. This growing separation forms the core of the model-dependent ambiguity that future colliders will face. Consequently, while a comprehensive, exhaustive multi-parameter sensitivity scan across all five saturation models is reserved for the companion paper \citep{ZardoBecker2026b}, this baseline comparison confirms that current LEP data lie comfortably within the predicted theoretical envelope, thereby validating the robustness of the benchmark established in this work.}

\section{Conclusions}
This paper has laid the theoretical groundwork for two-photon interactions at high energies. We have revisited the composite nature of the photon, the low‑energy QPM and VDM mechanisms, and the LEP evidence that forced the inclusion of the gluon component. The dipole formalism was introduced, and we have shown that the two phenomenological prescriptions for the dipole–dipole cross section (TKM and IKT) arise from two fundamentally different ways of mapping the interaction of two extended color dipoles onto the dipole–proton amplitude $\mathcal{N}(r,Y)$. 

A key conceptual point is that neither prescription is derived from QCD; both are heuristic ansätze. They break the exchange symmetry when the two dipoles have comparable sizes, and they rely on the untested assumption that a dipole scatters off another dipole as it would off a proton. By presenting a detailed description of the BK evolution equation and five saturation models (GBW, IIMS, rcBK, AGBS, WYKWC), we have provided a self‑contained reference for the theoretical tools that will be used in the companion paper~\citet{ZardoBecker2026b} to compute observables and to quantify the TKM–IKT ambiguity. 

While the full numerical evaluation of the TKM--IKT ambiguity is deferred to the companion paper, the present work provides the structural mapping required to interpret those quantitative results. Consequently, this study establishes a critical assessment of the state of the art, explicitly delineating where heuristic assumptions enter the formalism and where future first-principles developments will be required.

We emphasize that the present work does not claim to provide a first-principles derivation of the dipole–dipole cross section. Rather, it offers a systematic phenomenological benchmark that quantifies the current theoretical envelope. The TKM and IKT prescriptions are heuristic, their comparison is model-dependent, and their validity rests ultimately on experimental tests. Deviations of future cross-section measurements from the predicted theoretical envelope will mandate the formulation of a symmetric two-dipole evolution kernel and the introduction of a photon-specific saturation scale $Q_s^\gamma(x)$, constituting the foundational objective of the companion analysis.

\subsubsection*{Acknowledgment}
The author GZBw ould like to express special thanks to João Thiago de Santana Amaral for his extensive advice throughout the development of this work. His gratitude is extended to Yuri Kovchegov and Brandon Manley for their helpful discussions and valuable suggestions. This work was partially supported by the Brazilian funding agencies CNPq, CAPES, and INCT-FNA.

\bibliography{Zardo1AN}

\end{document}